\documentclass{iau}

\usepackage{amsmath}
\usepackage{graphicx}
\usepackage{multirow}
\usepackage{bm}

\usepackage{mathptmx}
\usepackage{newtxtext}
\usepackage[varvw]{newtxmath}

\usepackage{natbib}

\begin{document}

\lefttitle{Gizon et al.}
\righttitle{Solar Inertial Modes}

\jnlPage{1}{7}
\jnlDoiYr{2021}
\doival{10.1017/xxxxx}

\aopheadtitle{Proceedings IAU Symposium}
\editors{A. V. Getling \&  L. L. Kitchatinov, eds.}

\newcommand{\LG}[1]{\color{red} [ #1 ]\color{black}}
\newcommand{\JP}[1]{\color{cyan} [ #1 ]\color{black}}

\newcommand{\hanson}{\color{red}HFR(?) \color{black}}

\newcommand{\Omegaref}{{\Omega_0}}

\newcommand{\ii}{\textrm{i}}

\title{Solar Inertial Modes}

\author{Laurent~Gizon$^{1,2}$, Yuto~Bekki$^1$, Aaron~C.~Birch$^1$, Robert~H.~Cameron$^1$, Damien~Fournier$^1$, Jordan Philidet$^{1}$, B.~Lekshmi$^1$, Zhi-Chao~Liang$^1$}
\affiliation{$^1$Max-Planck-Institut f\"ur Sonnensystemforschung,  G\"ottingen, Germany}
\affiliation{$^2$Institut f\"ur Astrophysik und Geophysik, Georg-August-Universit\"at G\"ottingen, Germany\\ email: {\tt gizon@mps.mpg.de}}

\begin{abstract}
The Sun's global inertial modes are very sensitive to the solar differential rotation and to properties of the deep solar convection zone which are currently poorly constrained. These properties include the superadiabatic temperature gradient, the latitudinal entropy gradient, and the turbulent viscosity. The inertial modes also play a key role in controlling the Sun's large-scale structure and dynamics, in particular the solar differential rotation.  This paper summarizes recent observations and advances in the (linear and nonlinear) modeling of the solar inertial modes.  
\end{abstract}


\maketitle

\section{Introduction}

The modes of stellar oscillation are broadly classified into two categories \citep[e.g.,][]{Unno1989}. The \textit{spheroidal modes} of oscillation are associated with both horizontal and radial motions and include the acoustic and the gravity modes. In the presence of rotation, \textit{toroidal modes} of oscillation can also exist. The toroidal modes 
are associated with horizontal motion and are described by their radial vorticity eigenfunctions. The toroidal modes have frequencies in the inertial frequency range, i.e. of order the rotation frequency of the star. In the special case of uniform rotation, the toroidal modes are retrograde  Rossby modes restored by the Coriolis force.

The above classification is a simplification: the actual modes of solar oscillation in the inertial frequency range which have recently been observed are more complicated due to the Sun's differential rotation and, for example, the superadiabatic stratification of its convection zone. In this paper we present an overview of recent work on these hydrodynamic modes. Our aim is to outline progress in the observation, interpretation, and diagnostic potential of these modes. We also present preliminary observations of solar-cycle variations in the mode parameters.

The new field of \textit{inertial-mode helioseismology} has emerged out of three key observational papers:
(i) {\cite{Loeptien2018}} ,
(ii) {\cite{Gizon2021}} ,
and (iii)  {\cite{Hanson2022}}.  
The first paper reported the detection of equatorial (sectoral) Rossby modes on the Sun.
The second, the detection and identification of additional modes which owe their existence to the differential rotation, among which the $m=1$ high-latitude mode and other critical latitude modes. The third paper reported modes at mid latitudes with north-south antisymmetric vorticity.
The above papers are complemented by a number of  useful observational studies by \citet{Liang2019}, \citet{Proxauf2020}, \citet{Mandal2021},  to name a few.

Using numerical modeling, the solar inertial modes have been shown to be important not only to probe properties of the deep convection zone that are hardly accessible to the acoustic modes \citep[incl. superadiabaticity and turbulent diffusivities, see][]{Gizon2021}, but also in terms of their dynamical effect on the axisymmetric flows \citep{Bekki2024-Baroclinic}. Like acoustic modes, most inertial modes are linearly stable and stochastically excited by turbulent convection \citep{Philidet2023}. Some inertial modes, however, are linearly unstable and reach relatively large velocity amplitudes at high latitudes \citep{Fournier2022, Bekki2022-Linear}. 

\section{Observations of global modes of oscillation in the inertial frequency range}
\label{sec_obs}

\begin{figure}
    \centering
    \includegraphics[width=\textwidth]{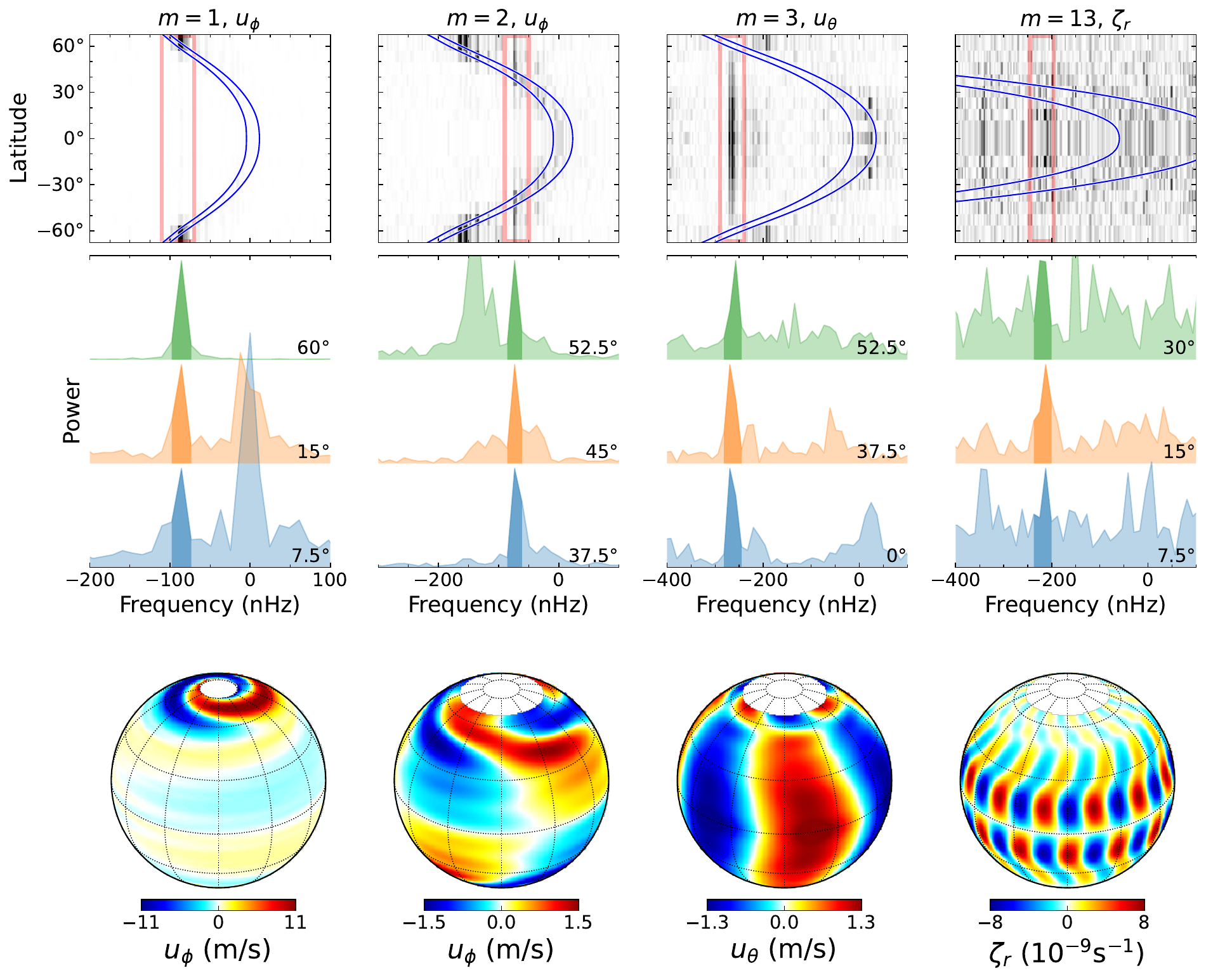}
    \caption{
    Example inertial modes observed in the power spectra of near-surface horizontal flow components at different latitudes, and the corresponding observed eigenfunctions. These modes are among many other modes  observed at any given value of $m$.    
    The four columns show (from left to right) the $m=1$ high-latitude mode at $-86$~nHz \citep{Gizon2021}, the $m=2$ mid-latitude mode at $-73$~nHz \citep{Gizon2021}, the $m=3$ equatorial (sectoral) Rossby mode at $-269$~nHz \citep{Loeptien2018}, and the $m=13$ high-frequency retrograde mode at $-220$~nHz \citep{Hanson2022}.
    The data were analysed in the Carrington frame of reference (a negative frequency indicates retrograde propagation in this frame). For clarity,
  the power spectra at fixed latitudes (middle row) have a reduced frequency resolution of $12$~nHz. The first three modes from the left have $u_\phi$  antisymmetric across the equator and $u_\theta$ is symmetric, while the fourth mode on the right has  antisymmetric radial vorticity $\zeta_r$. 
    }
    \label{fig:obs}
\end{figure}

Figure~\ref{fig:obs} shows low-frequency power spectra of solar oscillations for four particular values of the azimuthal wavenumber $m$. The power spectrum is computed for each colatitude $\theta$ according to
\begin{equation}
    P(\omega, \theta) = \left|  \sum_{\phi,t} u_j(\theta,\phi,t) e^{-\ii (m\phi - \omega t)} \right|^2 ,
\end{equation}
where $u_j$ is either the colatitudinal component of velocity ($u_\theta$) or the longitudinal component ($u_\phi$) inferred from ring-diagram analysis near the solar surface applied to 10 years of HMI observations.
The top row shows excess power corresponding to four selected modes, including the $m=1$ high-latitude mode identified by \citet{Gizon2021} and the $m=13$ mode reported by \citet{Hanson2022}.
As shown in the second row, the excess power associated with a mode is visible at the same frequency at multiple latitudes. We are thus in the presence of global modes of oscillation.
The last row of plots displays the eigenfunctions in either $u_\phi$, $u_\theta$ or the radial vorticity $\zeta_r$  extracted from narrow frequency windows centered on the mode frequencies. The typical mode lifetime is half a year (for a 20 nHz linewidth). Most of the modes that have been detected have  velocity amplitudes in the range $0.5$--$2$~m/s. The largest  amplitude mode is the $m=1$ high-latitude mode, with a velocity reaching   $\sim 15$~m/s on average and up to $20$~m/s during quiet Sun periods. The modes are quasi-toroidal, i.e. their motion is dominantly horizontal and the two components of  velocity can be described via a streamfunction.  The modes reported by \citet{Hanson2022} have low amplitudes of a few $0.1$~m/s.

\section{2D Linear analysis:  families of toroidal modes}

\begin{figure}
    \centering
    \includegraphics[width=\textwidth]{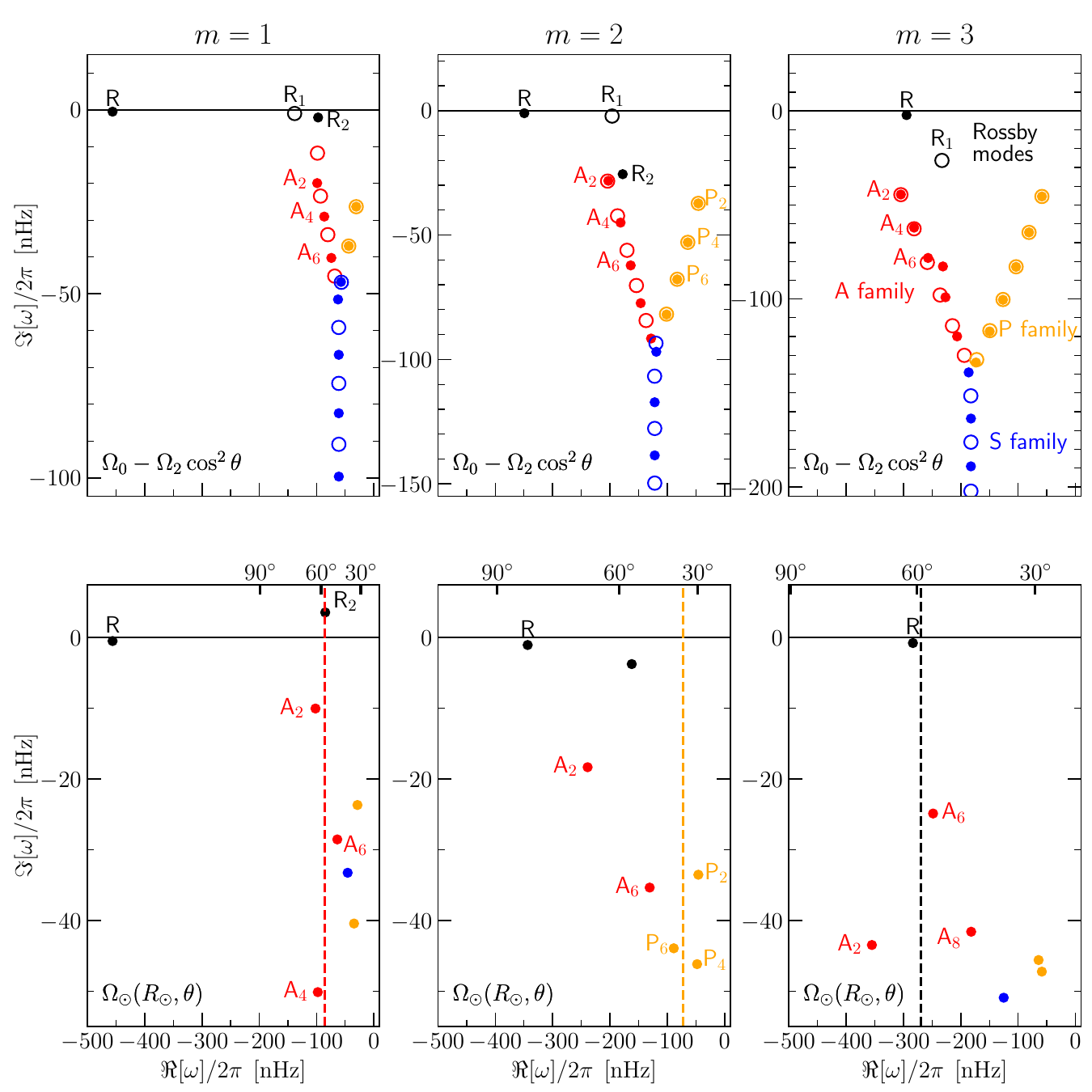}
    \caption{Eigenvalues in the complex plane for small-amplitude  toroidal  modes on a differentially rotating spherical surface \citep{Fournier2022}. Viscosity is specified via the Ekman number  $E=4 \times 10^{-4}$. The top panels show the real and imaginary parts of the eigenvalues for $m=1,2,3$ and a differential rotation profile $\Delta\Omega =  - \Omega_2\cos^2\theta$ with $\Omega_2/2\pi = -114$~nHz. Using the notations of \citet{Drazin2004}, the various families of hydrodynamic modes are labelled R for the Rossby modes (black), A for the wall modes (in red), S for the highly damped modes (blue), and P for the center modes (orange).  The full circles mark modes with radial vorticity that is symmetric across the equator (such as the first three modes shown in Fig.~\ref{fig:obs}), and  the open circles are for modes that are  antisymmetric.  The bottom panels show the eigenvalues of the symmetric modes for the differential rotation profile measured at the Sun's surface (the labels follow the tracks obtained 
    by slowly tracking the modes from the  top panel). The vertical dashed lines correspond to the observed  frequencies of the first three example modes from Fig.~\ref{fig:obs}. 
    }
    \label{fig:dispersion2d}
\end{figure}

Before looking at the full problem, it is useful to consider the much simpler problem of toroidal modes  on a spherical surface. Such modes are described in terms of their stream function
\begin{equation}
\Psi =  \psi(\theta) e^{\ii(m\phi-\omega t)} .
\label{eq.stream}
\end{equation}
We work in a frame that rotates with the Sun at the rate $\Omega_0$ (either the Carrington rotation rate or the equatorial rotation rate).
In this rotating frame, the governing equation  follows from the radial component of the curl of the linearized Navier-Stokes equation  \citep{Fournier2022}:
\begin{equation}
    \left(\omega- m\Delta\Omega \right)L_m  \psi  -m (2\Omegaref  - \Delta\Omega'')\psi 
    = \ii E\Omegaref L_m^2\psi, 
    \label{eqn:2d}
\end{equation}
where the quantity $\Delta\Omega (\theta)= \Omega(\theta)-\Omegaref$ is the latitudinal differential rotation measured in the rotating frame, 
$L_m$ is the Laplace-Beltrami operator on the unit sphere, $
\Delta\Omega''=\frac{1}{\sin\theta}\frac{d}{d\theta} \left( \frac{1}{\sin\theta} \frac{d}{d\theta}\left(\Delta\Omega\sin^2\theta\right) \right)  
$, and
$E = {\nu}/({r^2\Omegaref})$ is the Ekman number.
A nonzero $E$ accounts for the  turbulent viscosity $\nu$, which is an essential ingredient of  the problem.

\subsection{Uniform rotation}

In the special case of uniform rotation ($\Omega=\Omegaref$, $\Delta \Omega=0$), we obtain the classical Rossby modes with
\begin{eqnarray}
 \psi(\theta) = P_l^m(\cos\theta)
 \quad\textrm{and}\quad
\omega = - \frac{2m\Omegaref}{l(l+1)}   -   \ii   l(l+1)E \Omegaref , \label{eq:classical_rossby}
\end{eqnarray}
where the $P_l^m$ are associated Legendre polynomial of degree $l$ and azimuthal order $m$. The sectoral ($l=m$) modes are observed on the Sun  to have frequencies close to the model frequencies $\Re[ \omega] = - 2\Omega_0/(m+1)$ \citep{Loeptien2018}. The condition $E \lesssim 2\times 10^{-3}$ follows from the  observed   mode lifetimes \citep{Philidet2023},
implying  a turbulent viscosity $\nu \lesssim 2\times 10^{13}$~cm$^2$/s of order the mixing length estimate \citep{Munoz2011}. The observed eigenfunctions are  close  to $P_m^m(\cos\theta) \propto (\sin\theta)^m$, but with notable differences due to the presence of critical latitudes \citep{Gizon2020}, as we now discuss.

\subsection{Importance of differential rotation: critical latitudes}

Differential rotation means the rotation rate of the Sun is a function $\Omega(\theta, r)$ of colatitude $\theta$ and radius $r$. The factor $(\omega-m\Delta \Omega)$ in the first term of Eq.~(\ref{eqn:2d}) leads to the presence of critical latitudes $\theta_c$ where
\begin{equation}
\frac{\omega}{m}  = \Delta\Omega(\theta_c, r) \quad \textrm{in the corotating frame} .
\end{equation}
In the absence of viscosity, we have a singular eigenvalue problem  and the
eigenfunctions are not regular at the critical latitudes \citep[see, e.g.,][]{Balmforth1995}.
See also \citet{Watson1981} and \citet{Charbonneau1999} for a discussion of stability when $E=0$. The inclusion of viscosity  in Eq.~(\ref{eqn:2d})
leads to a regular eigenvalue problem of the 
Orr-Sommerfeld type (fourth-order eigenvalue problem). 
The top panels of Fig.~\ref{fig:dispersion2d} show the  corresponding eigenvalues for $E=4\times10^{-4}$ and in the case of a smooth differential rotation profile  $\Delta\Omega(\theta) =  - \Omega_2\cos^2\theta$ that approximates the Sun's angular velocity at low latitudes. For these parameters, all the modes are linearly stable and belong to three families of solutions in addition to the Rossby modes: the high-latitude modes, the critical-latitude (or center) modes, and the highly-damped modes, which are  analogous to the A, P and S families of  modes of a plane parabolic shear flow \citep[see][]{Drazin2004, Gizon2020}.
The lower panels of Fig.~\ref{fig:dispersion2d} show the modes when the observed surface differential rotation of the Sun is used instead of the two-term profile. Because the  solar rotation rate  drops sharply at high latitudes, an unstable mode appears in each of the $m=1$ and $m=2$ spectra.
The number of unstable modes as a function of the Ekman number was discussed by \citet{Fournier2022}.
The $m=1$ mode R$_2$  with a  streamfunction that resembles $P_3^1(\cos\theta)$ is unstable and has a frequency  very close to that of the  largest amplitude  mode in the observations (see Fig.~\ref{fig:obs}, left column). The high-latitude mode A$_2$ has a similar frequency but is linearly stable. Neither mode has an eigenfunction that matches the observed spiral pattern --- this requires 3D modeling and additional physics, as explained in the next section. The observed $m=2$ critical-latitude mode is in the range of frequencies corresponding to the P family \cite[also see figure~2 in][]{Gizon2021}. The frequencies and eigenfunctions of the equatorial  Rossby mode are satisfactorily reproduced in 2D, including their departure from sectoral spherical harmonics  \citep{Gizon2020, Fournier2022}. The 2D model however does not give the modes reported by \citet{Hanson2022} because these modes are not purely toroidal, see Sect.~\ref{sec.HFR}.

\section{Linear analysis in three dimensions} 
\label{sec.linear}

\subsection{Sectoral Rossby modes}
For the case of uniform rotation in three dimensions, the only modes that can easily be described analytically are the sectoral Rossby modes, which are purely toroidal. Their streamfunction separates, $\Psi = f(r) (\sin\theta)^m e^{\ii m\phi+2\ii\Omega t/(m+1)}$. Combining the radial and latitudinal components of the equation of motion, we obtain the radial dependence $f(r)\propto r^{m+1}$.  \citet{Proxauf2020} find that this model is not inconsistent with the radial dependence inferred from ring-diagram helioseismology, although only the top $8$~Mm could be studied with confidence. For a more sophisticated discussion of the sectoral Rossby modes, see \citet{Provost1981} and \citet{Damiani2020}. 

\subsection{Quasi-toroidal modes of the solar convection zone}
\label{sec_linear3d}

The Sun has radial structure and its inertial  modes are not separable in terms of radius and latitude \citep[e.g.,][]{Baruteau2013}.   
To study the linear stability of the eigenmodes of oscillation in the convection zone, we consider the linearized hydrodynamic equations (conservation of mass, momentum, and energy).
Here we outline the linear model presented by \citet{Bekki2022-Linear}. Ignoring the meridional flow for simplicity, the governing equations in spherical-polar  coordinates ($r,\theta,\phi$) are
\begin{eqnarray}
    \frac{D \rho'}{Dt}  &=& -\nabla\cdot (\rho \bm{u}'), \label{eq_lin:mass} \\
    \rho \frac{D \bm{u}'}{Dt} &=& -\nabla p' + \rho' \bm{g} +2\rho \bm{u}' \times (\Omega \bm{e}_z) -\rho r \sin{\theta}  (\bm{u}' \cdot \nabla \Omega ) \bm{e}_\phi +\nabla \cdot \mathcal{D}, \label{eq_lin:motion} \\
     \frac{D s'}{Dt} &=&  c_{p} \frac{\delta}{H_p}u'_r -\frac{u'_\theta}{r}\frac{\partial s}{\partial \theta} + \frac{1}{\rho T}\nabla \cdot (\rho T \kappa \nabla s'), \label{eq_lin:entropy}
\end{eqnarray}
with the linearized equation of state 
\begin{eqnarray}
    \frac{p'}{p}=\gamma \frac{\rho'}{\rho}+\frac{s'}{c_v}.
\end{eqnarray}
The perturbations from the background are denoted by primes. A realistic solar background stratification is considered. The total derivative along the solar differential rotation is given by 
\begin{eqnarray}
    \frac{D}{Dt}=\frac{\partial}{\partial t}+\Delta \Omega \frac{\partial}{\partial\phi},
\end{eqnarray}
where $\Delta \Omega = \Omega(r,\theta) - \Omega_0$ is the rotational profile from global helioseismology \citep{Larson2018}.
The superadiabaticity $\delta=-(H_p/c_p)\partial s/\partial r$ describes the small deviations from the adiabatic stratification.
The latitudinal entropy gradient is estimated under the assumption that the Sun's differential rotation is  in accordance with the thermal wind balance \citep{Rempel2005,Miesch2006},
\begin{eqnarray}
    \frac{\partial}{\partial\theta}\left( \frac{s}{c_{p}}\right) = \frac{r^2 \sin{\theta}}{g}\frac{\partial (\Omega^2)}{\partial z}. \label{eq:thermalwind}
\end{eqnarray}
The viscous stress tensor $\mathcal{D}$ is given by
\begin{eqnarray}
    \mathcal{D}_{ij} = \rho \nu \left[\left( \frac{\partial u'_i}{\partial x_j}+\frac{\partial u'_j}{\partial x_i}\right) -\frac{2}{3}(\nabla\cdot\bm{u}')\delta_{ij} \right].
\end{eqnarray}
Here, $\nu$ and $\kappa$ are the (turbulent) viscous and thermal diffusivities, which are assumed to be equal.

\begin{figure}
\centering
\includegraphics[width=0.8\textwidth, trim= 0 0 14.cm 0, clip]{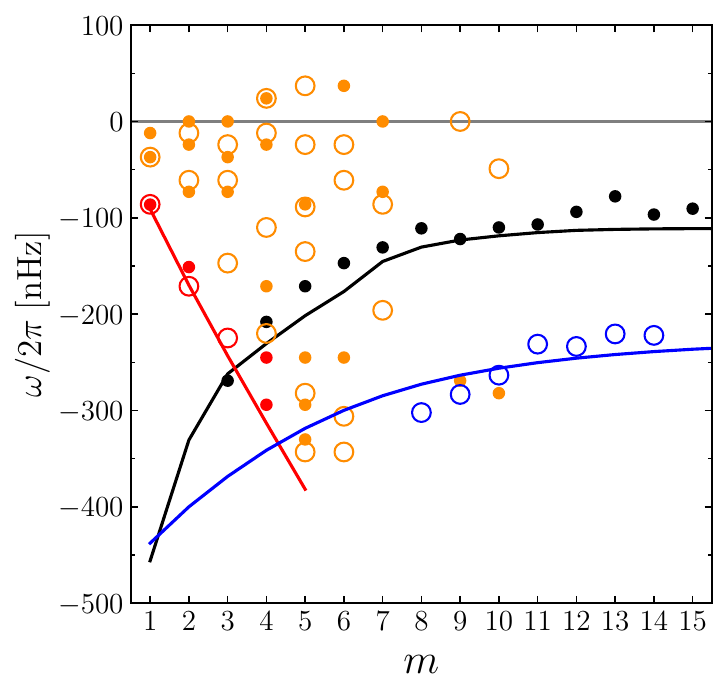}
    \caption{Comparison of the observed mode frequencies  in the Carrington frame (circles) and the real part of the mode frequencies from the 3D linear eigenvalue problem (curves). The filled circles mark modes with radial vorticity that is symmetric across the equator while the open circles are for modes that are antisymmetric.
    Black refers to the equatorial Rossby modes, red to the high-latitude modes, orange to the critical-latitude modes, and blue to the high-frequency retrograde mode.
    The observations are from \citet{Loeptien2018}, \citet{Gizon2021}, and \citet{Hanson2022}. The 
    model frequencies are from \citet{Bekki2022-Linear} and \citet{Bekki2024-HFR}.
    }
    \label{fig:dispersion}
\end{figure}

The above linearized equations (\ref{eq_lin:mass})--(\ref{eq_lin:entropy}) are combined into an eigenvalue problem assuming that the perturbations are proportional to $\exp{\ii(m\phi-\omega t)}$.
The computational domain extends from the base of the convection zone ($r=0.71R_{\odot}$) to slightly below the photosphere ($r=0.985R_{\odot}$).
At both radial boundaries, impenetrable and stress-free boundary conditions are used.
At the poles, all the perturbed quantities ($\rho'$, $\bm{u}'$, and $s'$) are set to be zero for azimuthal orders $m>1$.
For $m=1$, the special boundary condition $\partial u_{\theta}/\partial\theta=0$ is used to allow for the existence of a pole-crossing flow.
We numerically solve the eigenvalue equation using a second-order finite difference method.

The modes with   eigenfrequencies in the inertial frequency range $|\Re(\omega)| \lesssim 2\Omegaref$ 
 and with low damping rates are plotted in Fig.~\ref{fig:dispersion}. We see that the main families of  inertial modes are identified. 
\citet{Gizon2021} concluded that it is essential to include the latitudinal entropy gradient $\partial s/ \partial\theta$ from Eq.~(\ref{eq:thermalwind}) in the model 
to reproduce the correct spiral patterns of the $u_\phi$ eigenfunctions of the high-latitude modes. This also requires the conditions  $\langle \delta \rangle < 2\times 10^{-7}$ and $\langle \nu \rangle \leq 10^{12} $~cm$^2$/s on average in the convection zone. Both these upper limits  are an order of magnitude below the mixing length estimates. See \citet{Dey_thisvolume} for an analysis of the value of $\delta$ in the lower half of the convection zone.

\section{Excitation and damping mechanisms}

Inertial modes are present in  numerical simulations of rotating convection by \citet{Bekki2022-Amplitudes}, \citet{Matilsky2022}, and \citet{Blume2023-arXiv}. These simulations indicate that the  Rossby modes are excited to significant amplitudes in these self-consistent simulations. The high-latitude modes are not present in these simulation because the latitudinal differential rotation is not strong enough.
In the following paragraphs we discuss the excitation of the modes under specific setups.

\subsection{Stochastic excitation of equatorial Rossby modes}

According to the studies of Sect.~\ref{sec.linear}, many solar inertial modes are  linearly stable, including the  equatorial Rossby modes. The observed amplitudes of these modes are the result of a balance between the stochastic excitation of the waves by turbulent convection and their damping by turbulent viscosity.  
\citet{Philidet2023} added a  source term on the RHS of Eq.~(\ref{eq.stream}) to account for the  fluctuations of the divergence of the Reynolds stress tensor of the turbulence. As summarized in this volume \citep{Philidet_thisvolume}, this excitation mechanism is similar to the one used for   p~modes \citep{Goldreich1977}, except that the vorticity in the turbulence plays a major role here. Treating the turbulent flow as a known input, based on solar surface measurements \citep{Langfellner2015}, this model reproduces the observed amplitudes of the solar equatorial Rossby modes of $\sim 1$~m/s. It should be noted that there is a qualitative transition in the shape of the power spectrum between $m \lesssim 5$ where the inertial modes are clearly resolved in frequency, and $m \gtrsim 5$ where the modes overlap. This greatly complicates the interpretation of the observation, and implies that a model for the whole shape of the power spectrum is necessary to exploit the full potential of the solar inertial modes.

\subsection{Low-$m$ high-latitude modes are baroclinically unstable}

Using the eigenvalue solver  of Sect.~\ref{sec_linear3d}, \citet{Bekki2022-Linear} found that the low-$m$ high-latitude modes are very sensitive to the latitudinal entropy gradient $\partial s / \partial \theta$  in the  convection zone (or the corresponding latitudinal temperature gradient $\partial T / \partial \theta$).
Although the high-latitude modes are stable in 3D in the absence of $\partial s / \partial \theta$, they become linearly unstable with the presence of $\partial s / \partial \theta$ even when the radial stratification is adiabatic or subadiabatic.
Therefore, they are \textit{baroclinically} unstable.
This explains why the high-latitude modes have larger velocity amplitudes than the other inertial modes on the Sun \citep{Gizon2021}.

We note that, with a moderate value of turbulent viscosity ($\nu = 10^{12}$~cm$^{2}$/s), a very small latitudinal temperature variation of $\Delta T=T_{\rm pole}-T_{\rm eq} \approx 3$~K is enough to make the low-$m$ high-latitude modes unstable.
A realistic latitudinal entropy gradient can be estimated from Eq.~(\ref{eq:thermalwind}), which implies $\Delta T \approx 7$~K in the middle convection zone.
Once this  profile of $\partial s / \partial \theta$ is included, the linear model can successfully reproduce the observed   eigenfunctions of the high-latitude modes with $1 \leq m <  5$ (Fig.~\ref{fig:obs}, left panels), as well as their observed frequencies (Fig.~\ref{fig:dispersion}).

\section{Baroclinally unstable modes control differential rotation}

\subsection{3D nonlinear mean-field simulations}

To study the amplitudes of the high-latitude inertial modes, \citet{Bekki2024-Baroclinic}  carried out a series of hydrodynamic simulations of the large-scale flows in the Sun using a mean-field approach, where the small-scale convective motions are not explicitly solved for and the turbulent transport effects (incl. the $\Lambda$ effect) are parameterized. A parameter survey was conducted with varying baroclinic forcings whereby a negative (positive) latitudinal entropy gradient $\partial s / \partial \theta$ is generated in the northern (southern) hemisphere.

\subsection{Dynamical effect of modes on differential rotation}

\begin{figure}[]
    \centering
    \includegraphics[width=\textwidth]{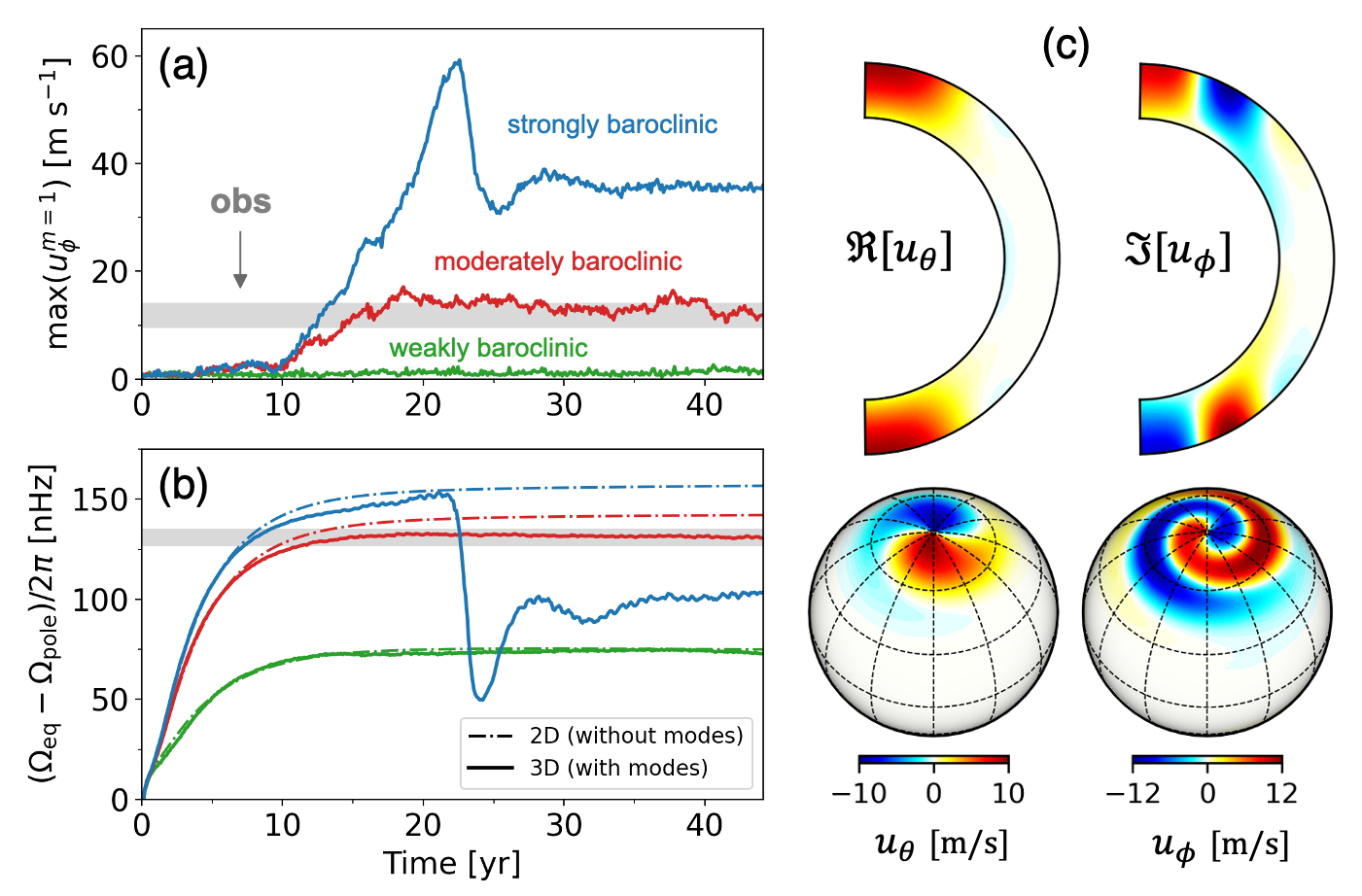}
    \caption{
    Results from 3D nonlinear simulations of the baroclinically-unstable modes \citep[][]{Bekki2024-Baroclinic}.
    (a) Amplitudes of the $m=1$ high-latitude mode at the surface as functions of time.
    Green, red, and blue curves show the results from the simulations with weak, moderate, and strong baroclinic forcing.
    The horizontal gray shade denote the observed value.
    (b) Latitudinal differential rotation, $\Omega_{\rm eq}-\Omega_{\rm pole}$, in the middle convection zone as functions of time.
    The solid and dashed curves denote the results from 3D full-spherical simulations (where the high-latitude modes are present) and from the 2D axisymmetric simulations (where the modes cannot exist), respectively.
    (c) Extracted eigenfunctions of horizontal velocities ($u_\theta$ and $u_\phi$) from the 3D nonlinear simulation (moderate case).
    Top and bottom rows show the meridional cuts and the surface eigenfunctions.
    }
    \label{fig:nonlinear3d}
\end{figure}

\citet{Bekki2024-Baroclinic} found that the amplitudes of the low-$m$ high-latitude modes are nonlinearly coupled to the latitudinal differential rotation in the convection zone.
Figures~\ref{fig:nonlinear3d}a and~b show the temporal evolution of the $m=1$ mode amplitude and the latitudinal differential rotation for three representative simulation cases where the baroclinic forcing is weak (green), moderate (red), and strong (blue).
When the baroclinicity is weak, the high-latitude modes are not excited to large amplitudes.
With a moderate amount of baroclinicity, the high-latitude modes become linearly unstable and the $m=1$ mode   velocity amplitude saturates at $\approx 13$~m~s$^{-1}$, which is consistent with the solar observation.
The extracted velocity eigenfunction of the $m=1$ mode from this case is shown in Fig.~\ref{fig:nonlinear3d}c.
The observed spiral feature in $u_\phi$ is nicely reproduced.
When the baroclinic forcing is even stronger, the modes are very efficiently baroclinically-excited and their amplitudes are larger than observed on the Sun.
We find that the high-amplitude modes play a significant dynamical role in limiting the latitudinal differential rotation via a nonlinear feedback.
The reduction of the latitudinal differential rotation is dominantly caused by the equatorward heat transport due to the high-latitude modes, which reduces the baroclinicity and changes the meridional circulation such that the poleward transport of angular momentum is increased.
This dynamical effect is not present in axisymmetric models where the non-axisymmetric inertial modes cannot exist (Fig.~\ref{fig:nonlinear3d}b).
We note that the horizontal Reynolds stresses associated with the high-latitude modes  imply a transport of angular momentum toward the equator \citep{Hathaway2013, Mandowara_thisvolume}, however this (direct) effect on differential rotation
is not as important as the (indirect) effect via heat transport.

\subsection{Latitudinal temperature gradient}

\citet{Bekki2024-Baroclinic} 
 find that the high-latitude modes place  an upper limit on  $\Delta T=T_{\rm pole}-T_{\rm eq}$. Using the 3D nonlinear mean-field simulations mentioned above, a relationship between $\Delta T$ and the amplitudes of the $m\leq 3$ high-latitude modes can be established. These simulations show that  the value of $\Delta T$ cannot exceed $7$~K. Interestingly, the observed mode amplitudes imply that this  maximum of $7$~K is likely reached in the Sun. This conclusion is also consistent with the value estimated from the thermal wind balance approximation, Eq.~(\ref{eq:thermalwind}).

\section{Modeling high-frequency retrograde modes}
\label{sec.HFR}

The retrograde modes with $l=m+1$ radial vorticity reported by \citet{Hanson2022} do not follow the dispersion relation of the classical Rossby modes given by Eq.~(\ref{eq:classical_rossby}), but have about three times the frequencies of the sectoral Rossby modes. \citet{Hanson2022} argued that these modes cannot be identified as  magneto-Rossby modes, coupled Rossby-gravity modes, nor  thermal Rossby modes.
The observations were interpreted  by \citet{Triana2022} as a particular class of inertial modes where substantial radial flows are involved in the equatorial regions. In this sense, they may be called \textit{non-toroidal} inertial modes. Although the linear model used by \citet{Triana2022} was highly simplified (e.g., assuming a fluid with constant density), this mode identification was later confirmed by \citet{Bhattacharya2023a} and \citet{Bekki2024-HFR} using more realistic linear models that include the solar-like background stratification.

Unlike the modes that are quasi-toroidal, non-toroidal inertial modes manifest a strong sensitivity to the background superadiabaticity $\delta=\nabla-\nabla_{\rm ad}$. 
Using the model described in Sect.~\ref{sec_linear3d},
\citet{Bekki2024-HFR} showed that the mode frequencies  are compatible with the observed frequencies only when the bulk convection zone (in the model) is weakly subadiabatic on average. 
{Several branches of enhanced power, with one branch corresponding to the high-frequency retrograde modes, have been reported in non-linear rotating convection simulations by \citet{Blume2023-arXiv}. In these simulations the branches have low quality factors. }

\section{Solar cycle variations of mode parameters}

\begin{figure}
    \centering
\includegraphics[width=0.9\textwidth]{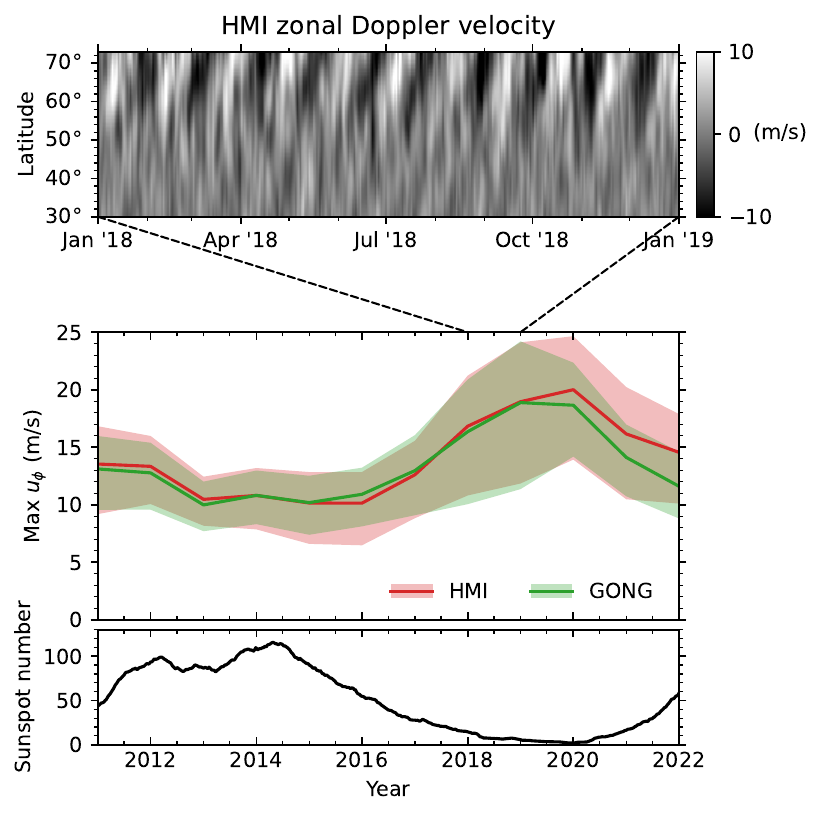}
    \caption{
    Map of zonal velocity (top) obtained from HMI line-of-sight Doppler signals during solar minimum showing the signature of the $m=1$ high-latitude mode whose amplitude varies with the solar cycle (bottom).
    The daily zonal velocity $V_{\rm zonal}$ were antisymmetrized across the equator and combined to form a synoptic map.
    The $m=1$ high-latitude mode of oscillation is seen as stripes above $50^\circ$ latitude with a period of about $34$~days as viewed from Earth.
    The mode amplitude averaged over the latitude range $60^\circ-75^\circ$ is estimated from $V_{\rm zonal}$ in a 3-yr sliding window. The mode amplitude is scaled by a  multiplicative factor of $2.9$ so that it is directly comparable to the deprojected longitudinal velocity $u_\phi$ shown in the bottom left panel of Fig.~\ref{fig:obs}.}
    \label{fig:HL1-cycle}
\end{figure}

Because of the close connection between the differential rotation profile and the characteristics of the  inertial modes, \citet{Goddard2020} suggested that the frequencies of the equatorial Rossby modes could vary with the sunspot cycle due to the temporal changes in rotation (varying zonal flows or `torsional oscillations'). The prediction was a positive correlation between the mode frequency shifts and the sunspot number. However, \citet{Waidele2023} and \citet{Lekshmi_thisvolume} measured an anticorrelation. This strongly suggests that the solar magnetic fields play the dominant role in the modulation of the mode properties.
\citet{Waidele2023} also noticed that the average power of the equatorial Rossby modes varies in phase with the sunspot cycle. \citet{Lekshmi_thisvolume} reach the same conclusion in this volume for the Rossby modes with $6\leq m \leq 10$.

Here we perform an analysis of the solar cycle dependence of the amplitude of the $m=1$ high-latitude mode using Dopplergrams as input. Because this mode is associated with horizontal flows in the near polar regions, it is detectable in the direct Doppler data.
The top panel of Fig.~\ref{fig:HL1-cycle} shows the zonal velocity computed from the HMI Dopplergrams during solar minimum, where the zonal velocity is given by \citep{Ulrich1988}
\begin{equation}
    V_{\rm zonal}(\theta) = \frac{ \sum_\phi  \Delta V_{\rm los}  \; \sin (\phi-\phi_{\rm c})}{\sum_\phi |\sin (\phi-\phi_{\rm c}) |}
    .
\end{equation}
In the above definition, $\Delta V_{\rm los}(\theta,\phi)$ is the Dopplergram corrected for the background signals and $\phi-\phi_{\rm c}$ is the longitude measured from the central meridian. The background signals consist of a  fit to the time-averaged limb shift, solar rotation, and meridional flows. In Fig.~\ref{fig:HL1-cycle}, $V_{\rm zonal}$ displays tilted stripes  at high latitudes \citep[these were already visible in the MWO data presented by][]{Ulrich1993,Ulrich2001}. 
The frequency of this pattern is measured to be $338$~nHz in the Earth frame, which corresponds exactly  to the frequency of the $m=1$ high-latitudes mode ($-86$~nHz in the Carrington frame).
The middle panel of Fig.~\ref{fig:HL1-cycle} shows the velocity amplitude $u_\phi$ of the  mode  estimated  from $V_{\rm zonal}$ over the latitude range $60^\circ$\,--\,$75^\circ$ as a function of time. 
The amplitude in $u_\phi$ varies between $10$ and $20$~m/s over this time period.
We find that the mode amplitude  is clearly anticorrelated with the sunspot number (bottom panel of Fig.~\ref{fig:HL1-cycle}). These new observations imply that a measurable  interaction happens between the $m=1$ mode and the Sun's global dynamo. Numerical modeling is needed to identify the nature of this interaction and to find out how and where it takes place in the convection zone.

\section{Outlook}

\subsection{Mode physics}

While helioseismology with acoustic modes is very largely based on the interpretation of the frequencies of the individual modes of oscillation using first-order perturbation theory, the interpretation of the inertial modes  cannot be separated from the study of  their amplitudes and eigenfunctions --- this requires more sophisticated forward modeling. In some regions of frequency space, the spectrum is too dense to be resolved: modes may overlap in frequency space and make the power spectrum difficult to decipher. Excitation and damping mechanisms  include the action of  turbulent convection, but also the nonlinear interaction of the modes with their environment. Solar inertial modes are not only important as new diagnostics of the solar convection zone, but also as regulating agents of  the global dynamics such as the Sun's meridional flow and differential rotation.

\begin{figure}[]
    \centering    
    \includegraphics[width=0.8\textwidth]{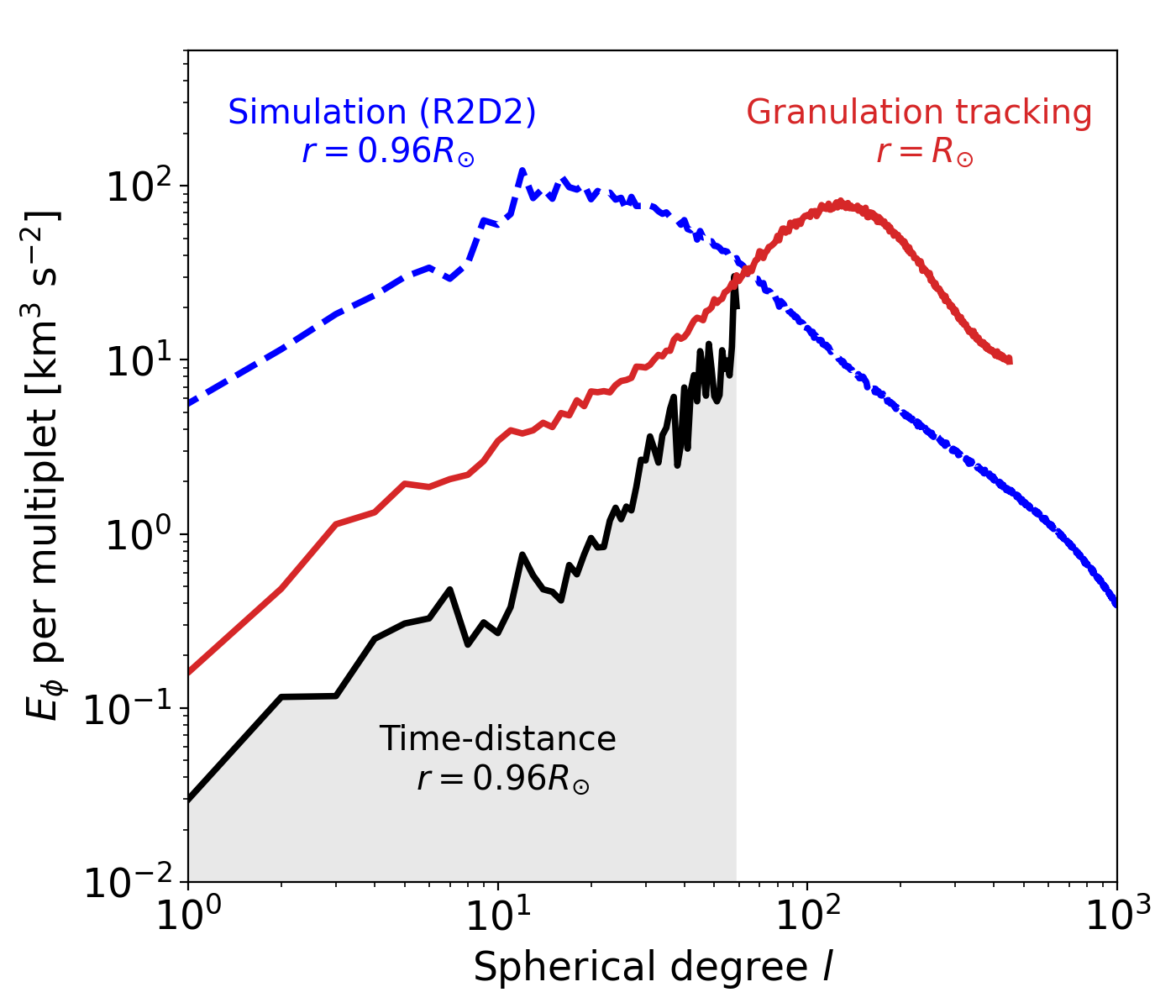}
    \caption{Power spectra of the longitudinal velocity near the solar surface.  
    The black solid curve shows the observational upper limit of $E_{\phi}(l)$ at $r=0.96R_{\odot}$ inferred by time-distance helioseismology \citep{Hanasoge2012} and revised by \citet{Proxauf_thesis}. The quantity $E_\phi$ represents the spatial power of the daily average of $u_\phi$, summed over all $m$ values at fixed angular degree $l$. 
    The red solid curve shows $E_{\phi}$ at the surface obtained by local correlation tracking of surface granulation \citep{Proxauf_thesis}. These observational data are available online \citep{Birch_edmond}. The blue dashed curve shows the spectrum from a numerical simulation of solar rotating magneto-convection \citep{Hotta2022}.}
    \label{fig:conundrum}
\end{figure}

\subsection{Opportunities to better understand the solar convection zone}


Figure~\ref{fig:conundrum} shows power spectra of the east-west velocity   obtained from a recent high-resolution numerical simulation by \cite{Hotta2022} and from two observational studies,  one from time-distance helioseismology  and another from tracking the granulation pattern in intensity \citep[see][and references therein]{Proxauf_thesis}. The order of magnitude discrepancy at low angular degrees (large horizontal  scales) between the numerical simulation and the observations is an outstanding problem \citep{Hanasoge2012,Gizon2012}. 

This problem is connected to the apparent lack of  north-south elongated convective modes in the solar surface observations. In numerical models these so-called banana cells are rotationally-constrained modes of rotating convection \citep[e.g.,][]{Gilman1977,Miesch2008}. The banana cells are robust features of the theory; they propagate prograde and can be understood in terms of the compressional $\beta$-effect or the conservation law of potential vorticity  \citep[e.g.,][]{Glatzmaier1981,Verhoeven2014}. They are a class of inertial modes sometimes called \textit{thermal Rossby modes}. Their properties have been studied in the linear regime \citep[][]{Glatzmaier1981,Bekki2022-Linear,Hindman2022,Jain2023} and in nonlinear rotating convection simulations \citep{Bekki2022-Amplitudes,Blume2023-arXiv}. Modes of mixed character sharing characteristics of prograde thermal Rossby and retrograde inertial modes have also been discussed by \citet{Bekki2022-Linear,Bekki2022-Amplitudes}, \citet{Jain2023}, and \citet{Blume2023-arXiv}. The impact of missing physics, such as, e.g., the importance of magnetic fields, should be assessed.

The inertial modes give us new information on the physical conditions deep inside the convection zone. In this summary paper we gave examples of what these modes are sensitive to, including the latitudinal temperature gradient, superadiabaticity, and the strength of the turbulent motions (through constraints on the turbulent viscosity) in the deep  convection zone. 
Looking further ahead, the inertial modes are likely to be affected by -- and to affect -- the magnetic field of the Sun. 
Modes of mixed character, gravito-magneto-inertial, also present novel opportunities and challenges to understand the tachocline \citep[see, e.g.,][]{Dikpati2018,Matilsky2022}.

\section*{Acknowledgements}
The Solar and Stellar Interiors Department of the Max Planck Institute for Solar System Research acknowledges generous support from the Max-Planck-Gesellschaft zur F\"orderung der Wissenschaften~e.V., the European Research Council  (ERC Synergy Grant WHOLE SUN 810218), and the Deutsche Forschungsgemeinschaft (SFB 1456/432680300 Mathematics of Experiment, project C04). Work on mode excitation by turbulence is supported by JP's Max Planck Partner Group at LESIA. The HMI data are courtesy of NASA/SDO and the HMI Science Team. This work utilizes data acquired by GONG instruments operated by NISP/NSO/AURA/NSF with contribution from NOAA. The sunspot numbers are from the World Data Center SILSO, Royal Observatory of Belgium, Brussels.


\end{document}